\documentclass[aps,prb,twocolumn,superscriptaddress]{revtex4}
\usepackage{amsmath}
\usepackage{graphicx}
\usepackage{amsfonts}
\usepackage{bbold}
\usepackage{color}
\usepackage{epstopdf}

\begin{document}

\title{Transport through asymmetric two-lead junctions of Luttinger liquid
wires}
\author{D.N. Aristov}
\affiliation{Petersburg Nuclear Physics Institute, Gatchina 188300, Russia}
\affiliation{Institut for Nanotechnology, Karlsruhe Institute of Technology, 76021
Karlsruhe, Germany }
\affiliation{Department of Physics, St.Petersburg State University, Ulianovskaya 1,
St.Petersburg 198504, Russia}
\author{P. W\"olfle}
\affiliation{Institut for Nanotechnology, Karlsruhe Institute of Technology, 76021
Karlsruhe, Germany }
\affiliation{Institute for Condensed Matter Theory, and Center for Functional
Nanostructures, Karlsruhe Institute of Technology, 76128 Karlsruhe, Germany}
\date{\today}

\begin{abstract}
We calculate the conductance of a system of two spinless Luttinger liquid
wires with different interaction strengths $g_{1},g_{2}$, connected through
a short junction, within the scattering state formalism. Following earlier
work we formulate the problem in current algebra language, and calculate the
scale dependent contribution to the conductance in perturbation theory
keeping the leading universal contributions to all orders in the interaction
strength. From that we derive a renormalization group (RG) equation for the
conductance. The analytical solution of the RG-equation is discussed in
dependence on $g_{1},g_{2}$. The regions of stability of the two fixed
points corresponding to conductance $G=0$ and $G=1,$respectively, are
determined.
\end{abstract}

\maketitle





\emph{Introduction}. Charge or spin transport in three-dimensional metallic
systems may be described in terms of Landau quasi-particles. In strictly
one-dimensional quantum wires quasi-particles are found to be unstable on
account of the interaction between electrons. An important part of that
physics is captured by the exactly solvable Tomonaga-Luttinger liquid (TLL)
model. \cite{GiamarchiBook,GoNeTs}
As is well known, transport through junctions of TLL wires is strongly
affected by the interaction in the wires, in some cases leading to a
complete blocking of transmission. The latter behavior can be traced back to
the formation of Friedel oscillations of the charge density around the
barrier, leading to an infinitely extended effective barrier potential in
the limit of low energy. The suitable language to describe this situation is
the renormalization group (RG) method, allowing to calculate the conductance
as a function of (length or energy) scale. Quite generally, the transport
behavior at low energy/temperature is dominated by only a few fixed points
of the RG flow. In the neighborhood of these fixed points the conductance is
found to obey power-law behavior as a function of temperature $T$ for the
infinite system or as a function of system length $L$ at zero temperature.
The problem of the two-wire junction has been studied first in the
pioneering works \cite{Kane1992,Furusaki1993}, using the method of
bosonization. Later, their results have been confirmed by many other authors,  see 
\cite{Safi1999} and references therein. For special values of the interaction exact results
have been obtained \cite{Weiss1995,Fendley1995}.

A purely fermionic formulation of the problem has been introduced in the
limit of weak interaction by Yue, Glazman and Matveev \cite{Yue1994}. We
have extended that theory to the regime of strong coupling by summing up an
infinite series of terms in perturbation theory, identified as the leading
and universal contributions \cite{Aristov2009}. The results obtained with
our method for the two-lead junction as well as the three lead junction with
time-reversal symmetry \cite{Aristov2010,Aristov2011a} and with magnetic
flux \cite{Aristov2011b} are in agreement with exact theoretical results,
where available. However, our results go beyond what has been obtained by
other authors in various ways. The majority of the previous works considered the
symmetric case of equal interaction strength in both half wires, with exception 
of the works \cite{Furusaki1996,Chamon1997} discussed below. 

In this paper we generalize our previous treatment \cite{Aristov2009} to the
case of two wires with different interaction strength $g_{1},g_{2}$. This
includes the case of a barrier at the end of a TLL wire, for which one of
the interaction parameters is zero, e.g.  $g_{2}=0$. As before, we confine
our considerations to spinless fermions.

\emph{The model}. To illustrate the system we are interested in, we first consider 
a tight binding Hamiltonian 
$\mathcal{H}_{tb}$ of free spin-less fermions describing two quantum wires
connected at a single junction by tunneling amplitudes: 
\begin{equation}
\mathcal{H}_{tb}=\Big[\sum_{j=1}^{2} 
\sum_{n=0}^{N}t_{0} c_{j,n}^{+}c_{j,n+1}+t_{b} c_{2,0}^{+}c_{1,0}\Big]+h.c.
\end{equation}%
Here $c_{j,n}^{+}$ creates a fermion in wire $j$ at site $n$, and $t_{b}$ is the
tunneling amplitude connecting the sites $(j,n=0)$ at the junction. 
The 2$\times $2 $S$-matrix characterizing the
scattering at this junction has the structure (up to overall phase factors
in the individual wires) 
\begin{equation}
S=%
\begin{pmatrix}
r & t \\ 
\widetilde{t} & r%
\end{pmatrix}%
=%
\begin{pmatrix}
\sin \theta  & i\cos \theta e^{-i\phi } \\ 
i\cos \theta e^{i\phi } & \sin \theta 
\end{pmatrix}
\label{S-parametrization}
\end{equation}%
We choose this parametrization in terms of the transmission and reflection
amplitudes $t,r$ , since it is readily generalizable to the case of
multi-wire junctions ($n$ wires, $n>2$ ). The above form of the $S$-matrix is completely general (up to irrelevant phase factors) and, in fact, defines our model.  Passing to the continuum limit,
linearizing the spectrum at the Fermi energy and adding forward scattering
interaction of strength $g_{j}$ in wire $j$ , we may write the TLL
Hamiltonian $\mathcal{H}$ in the representation of incoming and outgoing
waves as 
\begin{eqnarray}
\mathcal{H} &=&\int_{-\infty }^{0}dx\sum_{j=1}^{2}[H_{j}^{0}+H_{j}^{int} \Theta (-L<x<-l)]\,,
\\
H_{j}^{0} &=&v_{F}\psi _{j,in}^{\dagger }i\nabla \psi _{j,in}-v_{F}\psi
_{j,out}^{\dagger }i\nabla \psi _{j,out}\,, \\
H_{j}^{int} &=&2\pi v_{F}g_{j}\psi _{j,in}^{\dagger }\psi _{j,in}\psi
_{j,out}^{\dagger }\psi _{j,out}\,.
\end{eqnarray}%
We put $v_{F}=1$ from now on. The range of the interaction lies within the interval $(l,L)$, where $l>0$ serves as a ultraviolet cutoff and separates the domains of interaction and potential scattering on the junction; non-interacting leads correspond to large $|x|$ beyond $L$. 
 In terms of the doublet of incoming fermions $%
\Psi =(\psi _{1,in},\psi _{2,in})$ the outgoing fermion operators may be
expressed with the aid of the $S$-matrix as $\Psi (x)=S\cdot \Psi (-x)$ . We
express the interaction term of the Hamiltonian in terms of density
operators $\widehat{\rho }_{j,in}=\psi _{j,in}^{\dagger }\psi _{j,in}=\Psi
^{+}\rho _{j}\Psi =\widehat{\rho }_{j}$, and $\widehat{\rho }_{j,out}=\psi
_{j,out}^{\dagger }\psi _{j,out}=\Psi ^{+}\widetilde{\rho }_{j}\Psi =%
\widehat{\widetilde{\rho }}_{j}$\,, where $\widetilde{\rho }_{j}=S^{+}\cdot
\rho _{j}\cdot S$\,, as 
\begin{equation}
H_{j}^{int}=2\pi g_{1}\widehat{\rho }_{1}\widehat{\widetilde{\rho }}_{1}+
2\pi g_{2} \widehat{\rho }_{2}\widehat{\widetilde{\rho }}_{2}\,.
\end{equation}%
The matrices are given by $(\rho _{j})_{\alpha \beta }=\delta _{\alpha \beta
}\delta _{\alpha j}$ and $(\widetilde{\rho }_{j})_{\alpha \beta }=S_{\alpha
j}^{+}S_{j\beta }$. A convenient representation of 2$\times $2-matrices is
in terms of Pauli matrices $\sigma _{j}$, $j=1,2,3$, the generators of $%
SU(2)$ (see \cite{Aristov2009}). Notice that the interaction operator only
involves $\sigma _{3}$ (besides the unit operator $(\sigma _{0})_{\alpha
\beta }=\delta _{\alpha \beta }$). We note $Tr(\sigma _{j})=0$, $Tr(\sigma
_{j}\sigma _{k})=2\delta _{jk}$, $j=0,1,2,3$. Defining a two-component
vector $\mathbf{s}=(\sigma _{3},\sigma _{0})$, we have $\rho _{j}=\sqrt{1/2}%
\sum_{\mu }R_{j\mu }s_{\mu }$, where the $2\times 2$ matrix $\mathbf{R}$ has
the properties $\mathbf{R}^{-1}=\mathbf{R}^{T}$, $det\,\mathbf{R}=1$, and 
$R_{11}=R_{12}=-R_{21}=R_{22}=1/\sqrt{2}$. The outgoing amplitudes will be
expressed in terms of $\widetilde{\sigma }_{j}=S^{+}\cdot \sigma _{j}\cdot S$. 
With the aid of the $\sigma _{j}$ the $S$-matrix may be parametrized by
three angular (Euler) variables (see \cite{Aristov2009}), 
$S=e^{i\alpha_{1} \sigma_{3} /2} e^{i\alpha_{2} \sigma_{1} /2} e^{i\alpha_{3} \sigma_{3} /2}$. For the case under
consideration only two of these, $\theta, \phi$, are relevant: $S=
e^{-i\phi\sigma_{3} /2} e^{i(\pi - 2\theta)\sigma_{1} /2} e^{i\phi\sigma_{3} /2}$. The
corresponding elements of the $S$-matrix have been given in Eq.\ (\ref%
{S-parametrization}).

\emph{Parametrization of conductance matrix}. We may define a $2\times 2$
matrix of conductances $G_{jk}$ relating the current $I_{j}$ in lead $j$
(flowing towards the junction) to the electrical potential $V_{k}$ in lead $k
$ : $I_{j}=\sum_{k}G_{jk}V_{k}$. It follows from the conservation of charge
that $\sum_{j}G_{jk}=0$ and from invariance under a shift of the zero of
energy that $\sum_{k}G_{jk}=0$. Therefore, the conductance matrix has only
one independent element $G=(1-a)/2$, which relates the net current $I=\frac{1%
}{2}(I_{1}-I_{2})$ to the bias voltage $V=(V_{1}-V_{2})$, $I=GV$. We note
the relation $G=\frac{1}{2}(\mathbf{R}^{T}\mathbf{GR})_{11}$ , while all
other elements of $\mathbf{R}^{T}\mathbf{GR}$ are zero.

In the linear response regime the conductances are related to the $S$-matrix
by $G_{jk}=\delta _{jk}-Tr(\tilde{\rho }_{j}^{r}\rho _{k})=\delta
_{jk}-|S_{jk}^{r}|^{2}$, where the label $r$ indicates that the quantity is
fully renormalized by interactions. Defining $Y_{jk}=Tr(\tilde{\rho }%
_{j}^{r}\rho _{k})$ and using the above relation of $\tilde{\rho }%
_{j}^{r}$ and $\tilde{s}_{\mu }^{r}$, we see that the conductance
components may be expressed in terms of $\overline{Y}_{\mu \nu }=\frac{1}{2}%
Tr(\tilde{s}_{\mu }^{r}s_{\nu })$ as $Y_{jk}=(\mathbf{R}\overline{%
\mathbf{Y}}\mathbf{R}^{T}\mathbf{)}_{jk}$. Here and in the following bold
faced quantities marked with overbar are matrices in the transformed space
(indices $\mu ,\nu $ ). It follows from the analysis of the transformed
matrix $\mathbf{R}^{T}\mathbf{GR}$ given above that the matrix $\overline{%
\mathbf{Y}}$ has block structure, the nonzero elements being given by the
conductance parameter introduced above and by unity, $\overline{Y}_{11}=a$, $%
\overline{Y}_{22}=1$, $\overline{Y}_{12}=\overline{Y}_{21}=0$. From the
above relations we see that the parameter $a$ may be expressed by the angle $%
\theta $ in the above parametrization of the S-matrix as $a=-\cos2\theta $. We find therefore that $a$ is confined within the region $a\in
\lbrack -1,1]$.

\begin{figure*}[tbp]
\includegraphics*[width=1.6\columnwidth]{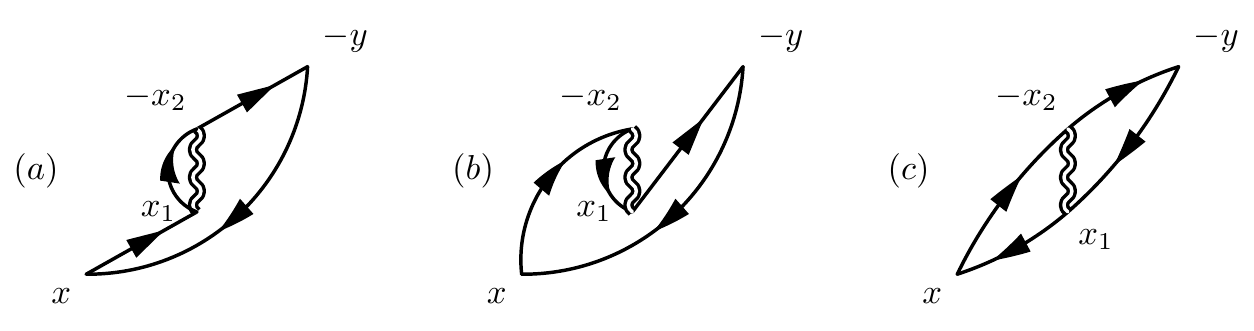}  
\caption{\label{fig:conduc} 
Feynman diagrams depicting the corrections to conductance. Two first diagrams, (a) and (b), correspond to  Eq.  (\protect\ref{LadContrib}), the third diagram (``vertex correction'') vanishes in the static limit, external $\Omega \to0$.  
}
\end{figure*}

\emph{Perturbation theory}. We now calculate the conductance in perturbation
theory in the interaction. In first order we have to evaluate the diagrams
depicted in Fig.\  \ref{fig:conduc}. Here solid lines denote Green's functions in
position-energy representation ($\omega _{n}$ are fermionic Matsubara
frequencies), $\mathcal{G}(x;\omega _{n})=-isign(\omega _{n})\theta (x\omega
_{n})e^{-x\omega _{n}}$\,. The double wavy lines denote the interaction operator,
which in the lowest order is 
given by the matrix $\overline{g}_{\mu \nu }=(\mathbf{R}^{T}\mathbf{gR)}%
_{\mu \nu }$, with $g_{jk}=\delta _{jk}g_{j},$ in the transformed or initial
representations, respectively. At the ends of the interaction lines
operators $\sigma _{3}$ and $\widetilde{\sigma }_{3}$ are attached depending
on whether $x<0$ or $x>0$. As a result, one finds in lowest order in the
interaction

\begin{equation}
\begin{aligned}
\overline{Y}_{\mu\nu }^{(1)} 
&=-\tfrac{1}{2}\mbox{Tr}\left( \widehat{\overline{W}}_{\mu\nu }\widehat{%
\overline{W}}_{\mu ^{\prime }\nu ^{\prime }}\right) \overline{g}_{\nu
^{\prime }\mu ^{\prime }}\Lambda 
\\ &=- \tfrac{1}{2}  \delta_{\mu1}\delta_{\nu1}(a^{2}-1)(g_{1}+g_{2})%
\Lambda \,.
\end{aligned}
\label{1stOrder}
\end{equation}%
Here the $\widehat{\overline{W}}_{\mu \nu }=(\mathbf{R}^{T}\widehat{\mathbf{W%
}}\mathbf{R)}_{\mu \nu }=\frac{1}{2}[s_{\mu },\widetilde{s}_{\nu }]$, and $%
\widehat{W}_{jk}=[\rho _{j},\widetilde{\rho }_{k}]$ , are 2$\times $2
matrices for each pair of $\mu \nu $ (or $jk$) and the trace operation $\mbox{Tr}$
is defined with respect to that matrix space, whereas the $\overline{g}_{\mu \nu
}=(\mathbf{R}^{T}\mathbf{gR)}_{\mu \nu }$, with $g_{jk}=\delta _{jk}g_{j}$ are
scalars. Notice that only one matrix element, 
$\widehat{\overline{W}}_{11}$, is nonzero.

\begin{figure}[tbp]
\includegraphics*[width=0.6\columnwidth]{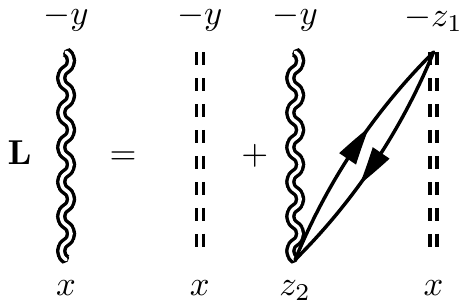}  
\includegraphics*[width=0.7\columnwidth]{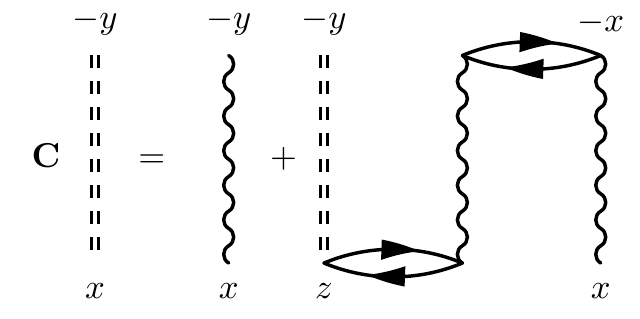}  
\caption{\label{fig:IntegEq} 
Feynman diagrams depicting the integral equations for the renormalized interaction, Eqs.(\ref{Linteq}) and (\ref{Cinteq}).    
}
\end{figure}

We may extend this analysis into the strong coupling regime by summing up
infinite classes of contributions in perturbation theory. It has been found
in \cite{Aristov2009} that the diagrams shown in Figs.\ \ref{fig:conduc} and 
\ref{fig:IntegEq} provide the leading terms in
the neighborhood of the fixed points. Their contribution is universal in
contrast to other higher order terms (see below). These may be interpreted as a
renormalization of the bare interaction, $ 2\pi g_{kl}\delta
(x-y)\rightarrow L_{kl}(x,y;\omega _{n})$. The contribution to conductance stemming from the first two diagrams (a) and (b) in Fig.\ \ref{fig:conduc}
in this ladder
approximation is given by
\begin{widetext}
\begin{equation}
\begin{aligned}
G^{(a+b)} &=-\frac{1}{4}T^{2}\sum_{\epsilon ,\omega }\int_{l}^{L}
dx_{1}dx_{2}\int_{L}^{\infty}dy\; \overline{L}_{11}(x_{1},x_{2};\omega ) 
\mathcal{G}(y+x;\epsilon +\Omega ) \\
&\times \left( \mathcal{G}(x_{1}-x;\epsilon )\mathcal{G}(-x_{1}-x_{2};\epsilon
-\omega ) \mathcal{G}(x_{2}-y;\epsilon )
+\cos4\theta\, \mathcal{G}(-x_{2}-x;\epsilon )\mathcal{G}(x_{1}+x_{2};\epsilon
+\omega ) \mathcal{G}(-x_{1}-y;\epsilon )
\right)
\,,  
\end{aligned}
\label{LadContrib}
\end{equation}%
Here only the $(1,1)$ element of $\overline{\mathbf{L}}$ enters,
corresponding to the fact that operators $\sigma _{3}$ and $\widetilde{%
\sigma }_{3}$ are attached to the ends of the renormalized interaction
line. The factor $\cos4\theta$ appears as 
$\text{Tr}(\sigma _{3}\widetilde{\sigma }_{3}\sigma _{3}\widetilde{\sigma }_{3})/2$. 
 In the limit of zero temperature we may convert the summation over
Matsubara frequencies to an integration along the imaginary axis. Another contribution
to $G$ is obtained by reverting the arrows on the fermion lines in  Fig.\ \ref{fig:conduc}a, Fig.\ \ref{fig:conduc}b, and doubles the above result. 
Performing
the integrations on $\epsilon $ and on $y$ from $L$ to $\infty $ and taking
the limit $\Omega \rightarrow 0$ we find 
\begin{equation*}
G^{(L)}= (1-a^{2})\int dx_{1}dx_{2}\frac{d\omega}{2\pi} \;\bar{L}%
_{11}(x_{1},x_{2};\omega )\theta (\omega )e^{-\omega (x_{1}+x_{2})}\,,
\end{equation*}%
It is useful to first calculate $\mathbf{L}$ in the initial representation,
where the interaction matrix $\mathbf{g}$ is diagonal, but the matrix $%
\mathbf{Y}$ is nondiagonal. Then $\mathbf{L}$ is found to satisfy the
following linear integral equation for $\omega>0$ (Ref.\ [\onlinecite{Aristov2009}], note that the definitions  of $\mathbf{Y}$ and for the strength of interaction there are different):

\begin{equation*}
\begin{pmatrix} \mathbf{L}(x,y;\omega ) \\   \mathbf{L}_{2}(x,y;\omega )
\end{pmatrix}
=2\pi \mathbf{g}\delta (x-y)
\begin{pmatrix}1 \\ 0 \end{pmatrix}
- 2\pi 
\int_{l}^{L}dz 
\begin{pmatrix} \mathbf{g}   \mathbf{Y} \Pi(x+z,\omega),& 
\mathbf{g}   \Pi(x-z,\omega) \\ 
\mathbf{g}   \Pi(z-x,\omega), & 0   \end{pmatrix}
\begin{pmatrix} \mathbf{L}(z,y;\omega ) \\   \mathbf{L}_{2}(z,y;\omega )
\end{pmatrix} 
\end{equation*}%
with the fermionic loop $\Pi(x,\omega_{n}) = (2\pi)^{-1} (\delta(x)-|\omega_{n}| 
\theta(x\omega_{n}) e^{-x\omega _{n}})$. Expressing $ \mathbf{L}_{2}(x,y;\omega )$ via 
$ \mathbf{L}(x,y;\omega )$ we have

\begin{equation*}
\mathbf{L}(x,y;\omega )=2\pi  \widetilde{\mathbf{g}}\delta (x-y)+\omega \widetilde 
{\mathbf{g}}
\int_{l}^{L}dz \left[(\mathbf{Y}-\tfrac{1}{2}\mathbf{g})e^{-\omega (x+z)}-\tfrac{1}{2%
}\mathbf{g}e^{-\omega |x-z|} \right]\mathbf{L}(z,y;\omega ),
\end{equation*}%
with $\widetilde g_{j} = g_{j} / d_{j}^{2}$ and $d_{j}= \sqrt{1-g_{j}^{2}}$. 
In order to solve this integral equation we first calculate a partial
summation, defined by \bigskip $\mathbf{C}(x,y;\omega )=\lim_{Y\rightarrow 0}%
\mathbf{L}(x,y;\omega )$. Then $\mathbf{L}(x,y;\omega )$ will be the
solution of
\begin{equation}
\mathbf{L}(x,y;\omega )=\bigskip \mathbf{C}(x,y;\omega )+\frac{\omega }{2\pi 
}\int_{l}^{L}dz_{1}dz_{2}\mathbf{C}(x,z_{1};\omega )\mathbf{Y}e^{-\omega
(z_{1}+z_{2})}\mathbf{L}(z_{2},y;\omega )\,,
\label{Linteq}
\end{equation}
and $\bigskip \mathbf{C}(x,y;\omega )$ satisfies the integral equation
\begin{equation}
\mathbf{C}(x,y;\omega )=2\pi \widetilde{\mathbf{g}}\delta (x-y)-\tfrac{1}{2}\omega 
\widetilde{\mathbf{g}}\mathbf{g} \int_{l}^{L}dz \left[e^{-\omega (x+z)}+e^{-\omega |x-z|} \right]\mathbf{C}%
(z,y;\omega ) \,.
\label{Cinteq}
\end{equation}
These integral equations are shown diagrammatically in Fig.\ \ref{fig:IntegEq}. 

We now define $\mathbf{C}(x,y;\omega )=2\pi \widetilde{\mathbf{g}}\delta (x-y)-\mathbf{C%
}_{1}(x,y;\omega )$, so that the inhomogeneity in the integral equation for $%
\mathbf{C}_{1}(x,y;\omega )$ is differentiable

\begin{equation*}
\mathbf{C}_{1}(x,y;\omega )=\pi \omega \widetilde{\mathbf{g}}^{2}\mathbf{g} \left[e^{-\omega
(x+y)}+e^{-\omega |x-y|} \right]-\tfrac{1}{2}\omega \widetilde{\mathbf{g}}\mathbf{g}
\int_{l}^{L}dz \left[e^{-\omega (x+z)}+e^{-\omega |x-z|} \right]\mathbf{C}_{1}(z,y;\omega
)  \,,
\end{equation*}%
The integral equation for $\mathbf{C}_{1}(x,y;\omega )$
may be converted into a second order differential equation

\end{widetext}

\begin{equation*}
\left[ \frac{\partial ^{2}}{\partial x^{2}}\mathbf{1}-\omega ^{2}(\mathbf{1}%
+\widetilde{\mathbf{g}}\mathbf{g})
\right]\mathbf{C}_{1}(x,y;\omega )=-2\pi \omega ^{2} \widetilde{\mathbf{g}}^{2}\mathbf{g}
\delta (x-y),
\end{equation*}%
Since the matrix $\mathbf{g}$ is diagonal, $\mathbf{C}_{1}$ is diagonal and
we have two uncoupled differential equations for the components $C_{1,j}$, $%
j=1,2$\,. Taking into account 
the boundedness of $C_{1,j}$, the general solution is given by  
$C_{1,j}(x,y;\omega )= \pi \omega _{j} \widetilde g_{j} g_{j}^{2} ( A_{j}(y)e^{-\omega _{j} x} + e^{-\omega _{j}|x-y|} )$ , where $\omega
_{j}=\omega/ d_{j}$. It follows from the
boundary conditions at $x=0$ that $A_{j}(y)= e^{-\omega _{j} y}$. 
The quantity $\mathbf{C}$ is thus a diagonal matrix given by

\begin{equation*}
\begin{aligned}
\mathbf{C}(x,y;\omega ) & =2\pi \mathbf{d}^{-2}{\mathbf{g}}\delta (x-y)
\\ & -\pi \omega \mathbf{d}^{-3}\mathbf{g}^{3}
\left[e^{-\omega(x+y)/ \mathbf{d}}+e^{-\omega |x-y|/ \mathbf{d}}\right]
\end{aligned}
\end{equation*}%

We now return to the integral equation for $\mathbf{L}(x,y;\omega)$. Since
the kernel is separable it is easily solved. We define the auxiliary matrix
functions $\mathbf{U}(y;\omega )=\int dxe^{-\omega x}\mathbf{L}(x,y;\omega )$, $\mathbf{V}(y;\omega )=\int dxe^{-\omega x}\mathbf{C}(x,y;\omega )=2\pi 
 \mathbf{d}^{-1} \mathbf{g}e^{-\omega y/ \mathbf{d}}$ as well as the matrix $\mathbf{Q}^{-1}= \frac{\omega }{2\pi }
 \int dxdye^{-\omega (x+y)}\mathbf{C}(x,y;\omega )=%
\mathbf{g}(\mathbf{1}+\mathbf{d})^{-1}$ . Multiplying the above integral
equation (\ref{Linteq}) by $e^{-\omega x}$ and integrating over $x$ we find $\mathbf{U}=%
\mathbf{V}+\mathbf{Q}^{-1}\mathbf{YU}=(\mathbf{1}-\mathbf{Q}^{-1}\mathbf{Y}%
)^{-1}\mathbf{V}$ and we finally get

\begin{equation*}
\begin{aligned}
\mathbf{L}(x,y;\omega ) &=\bigskip \mathbf{C}(x,y;\omega )
\\ &+\frac{\omega }{2\pi 
}\mathbf{V}(x;\omega )\mathbf{Y}(\mathbf{1}-\mathbf{Q}^{-1}\mathbf{Y})^{-1}%
\mathbf{V}(y;\omega )
\end{aligned}
\end{equation*}%
This result is now substituted into the expression for the conductance.
Performing the integration over $y$ first we get $\int dye^{-\omega y}\mathbf{L%
}(x,y;\omega )=\mathbf{V}(x;\omega )[\mathbf{1}+\mathbf{Y}(\mathbf{1}-%
\mathbf{Q}^{-1}\mathbf{Y})^{-1}\mathbf{Q}^{-1}]=\mathbf{V}(x;\omega )[
\mathbf{1}-\mathbf{YQ}^{-1}]^{-1}$ . Next integrating over positive $\omega $
yields $\int (d\omega /2\pi )e^{-\omega x}\mathbf{V}(x;\omega )[\mathbf{1}-%
\mathbf{YQ}^{-1}]^{-1}=(1/x)\mathbf{g}(\mathbf{1}+\mathbf{d})^{-1}[%
\mathbf{1}-\mathbf{YQ}^{-1}]^{-1}= (1/x)[\mathbf{Q}-\mathbf{Y}]^{-1}$. 
Finally, the integration over $x$ produces the scale dependent logarithm $%
\Lambda =\ln (L/l)$ \ . The conductance in the ladder approximation is found
after transforming to the rotated basis (quantities with overbar) and taking
the $(1,1)$-element of the matrix,

\begin{equation*}
G^{(L)}=\tfrac{1}{2}(1-a)-(1-a^{2})\left[(\overline{\mathbf{Q}}-\overline{\mathbf{Y%
}})^{-1}\right]_{11}\Lambda 
\end{equation*}

\emph{Renormalization group equations}. The renormalization of the
conductances by the interaction is determined from the scale dependent
contributions in perturbation theory. Differentiating these results with
respect to $\Lambda $ (and then putting $\Lambda =0$) we find the RG
equation for the quantity $a=1-2G$ in the ladder approximation 
\begin{equation}
\frac{da}{d\Lambda }=2(1-a^{2})\left[(\overline{\mathbf{Q}}-\overline{\mathbf{Y}}%
)^{-1}\right]_{11}  \label{eq:main}
\end{equation}%
Here $\overline{\mathbf{Y}}=diag(a,1)$ and $\overline{\mathbf{Q}}=\mathbf{R}%
^{T}\mathbf{QR}$ , such that $\overline{Q}_{11}=\overline{Q}_{22}=\overline{Q%
}_{+}=(q_{1}+q_{2})/2$ and $\overline{Q}_{12}=\overline{Q}_{21}=\overline{Q}%
_{-}=(q_{1}-q_{2})/2$ where $q_{j}=(1+\sqrt{1-g_{j}^{2}}%
)/g_{j}=(1+K_{j})/(1-K_{j})$, and $K_{j} = \sqrt{(1-g_{j})/(1+g_{j})}$
 is the usual Luttinger liquid
parameter for wire $j$. We define 
\begin{equation}
 \gamma = 
\overline{Q}_{+}- {\overline{Q}_{-}^{2} }/( {\overline{Q}_{+}-1})= 
\frac{K_{1}^{-1}+K_{2}^{-1}+2}{K_{1}^{-1}+K_{2}^{-1}-2}, 
\label{gamma}
\end{equation}
(note that $|\gamma|>1$ for any $K_{1,2} >0$ ), then 
the RG-equation takes the explicit form
\begin{equation}
\frac{da}{d\Lambda }=2\frac{a^{2}-1}{a-\gamma } \,. 
\end{equation}

\begin{figure}[tbp]
\includegraphics*[width=0.8\columnwidth]{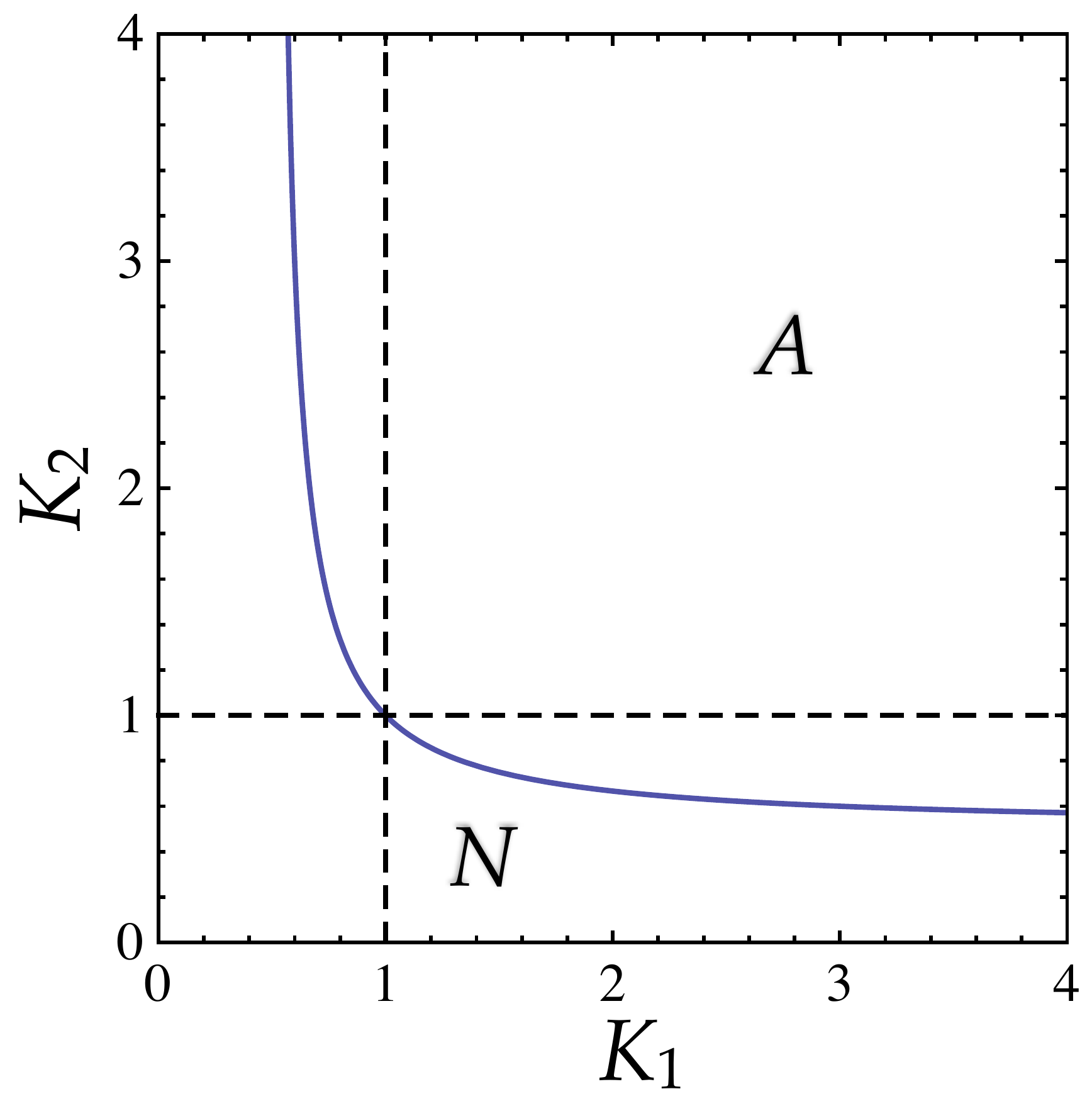}   
\caption{\label{fig:K1K2} 
Stability regions, labeled by the corresponding fixed point are shown in $K_{1}$-$K_{2}$ plane. The hyperbola separating the regions is shown by solid line, the non-interacting values $K_{1,2}=1$ are shown by dashed lines.  
}
\end{figure}

\emph{RG flow and conductance}. The fixed points of the above RG equation
are labelled $N$ , at  $a=1$, $G=0$,  (complete separation of the wires)
and $A$ at $a=-1$, $G=1$ (ideal conductance through the junction). In order
to discuss the stability and to calculate the conductance we rewrite the
RG-equation in terms of the conductance

\begin{equation}
\frac{dG}{d\Lambda }=-4\frac{G(1-G)}{2G-1+\gamma }=\beta (G)
\end{equation}%
Stability of a fixed point requires that the derivative of $\beta (G)$ at
the fixed point is negative. At fixed point $N$ this translates into  $%
\gamma -1>0$, or more explicitly, $q_{1}+q_{2}>2$, and in terms of the
Luttinger parameters $K_{1}^{-1}+K_{2}^{-1} > 2$
i.e. either both interactions should be repulsive, or one of them is\
attractive, but weak relative to the repulsive one. At fixed point $A$ the
condition is  $\gamma +1<0$, or $q_{1}+q_{2}<2,$ and therefore 
$K_{1}^{-1}+K_{2}^{-1} <2$ , meaning that the
interaction is predominantly attractive, but a weaker repulsive interaction
in one of the wires is possible. The line separating the stability regions
in the $K_{1}$-$K_{2}$-plane is the hyperbola, $(K_{1}-\frac12)(K_{2}-\frac12)=\frac14$, passing through the (no interaction) point $K_{1}=K_{2}=1$. This hyperbola and  corresponding stability regions are shown in Fig.\ \ref{fig:K1K2}

The RG-equation may be integrated to give 

\begin{equation}
G^{1-\gamma }(1-G)^{1+\gamma }=c(L/l) ^{4}
\label{eq:solution}
\end{equation}%
In the vicinity of fixed points $N$ and $A$ , respectively, we then find the
power laws

\begin{eqnarray}
G &=&c_{N}(l/L) ^{4/(\gamma -1)}, \qquad \text{at   }N \\
G &=&1-c_{A}(L/l) ^{4/(\gamma +1)}, \quad \text{at   }A
\end{eqnarray}
Explicitly we have $4/(\gamma -1) =  (K_{1}^{-1}+K_{2}^{-1}-2)$,   and 
$4/(\gamma +1) = 2(K_{1} + K_{2}-2K_{1}K_{2})/(K_{1}+K_{2})$. At $K_{1}=K_{2}=K$ we recover the well-known \cite{Kane1992,Furusaki1993} exponents $2(K^{-1}-1)$, $2(1-K)$.

Of special interest is the case $g_{1}=g$, \ $g_{2}=0$, when we have $\gamma =2q-1=(1+3K)/(1-K)$ and the power laws
are given by  $G=c_{N}(L/l) ^{-(K^{-1}-1)}$ at $N$ and $1-G=c_{A}(L/l)
^{-2(K-1)/(K+1)}$ at  $A$. We may compare these findings  with results obtained in 
Ref.\  [\onlinecite{Furusaki1996}] for a non-symmetric position of an impurity in a TLL wire. 
When taking the impurity position sufficiently close to the interface between the TLL wire and a non-interacting lead,  scaling exponents identical to the above ones were found in 
  \cite{Furusaki1996}, as can be read off  from Eqs.\  (10) and (13) there.

Thus we recover the correct exponents in the scaling law for the conductance. We obtained them by summation of the ladder sequence for the renormalized interaction in the presence of the junction. As was shown in \cite{Aristov2009},  the first contributions to conductance beyond the ladder series appear in the third order of interaction and are of three-loop type (the two-loop RG contributions are absent). These latter terms do not contribute to above scaling exponents, but define the relation between the prefactors $c_{N}$ and  $c_{A}$. The Eq.\ (\ref{eq:solution}) implies that both $c_{N}$ and $c_{A}$ depend on the initial conditions, encoded in the amplitude $c$. It is easily seen, that the ladder approximation corresponds  to the relation 
\[ c_{N}^{1-\gamma} / c_{A}^{1+\gamma} = 1 \,,\] 
It was shown \cite{Aristov2009}  that the three-loop corrections change this ratio to values, different from unity and depending on the strength of interaction and regularization scheme, i.e. non-universal.

Our expression (\ref{gamma}) shows that the boundary exponents 
depend only on the sum $K_{1}^{-1} + K_{2}^{-1}$. This is in precise 
agreement with Eq.\ (13) in  Ref.\ \cite{Chamon1997}, where  the scaling exponent of the point contact between two chiral (Hall edge) states was derived.  On the basis of this observation it was suggested there, that the combination of two chiral states, $K_1=1/3$ and $K_2 =1$, 
is equivalent to the well-known problem 
of impurity in TLL wire with $K_1=K_2 =1/2$, which can be fully solved. \cite{Kane1992}
We note here that the coincidence of the scaling exponents at two fixed points $N$ and 
$A$ might not necessarily mean the coincidence of 
the full scaling form for the conductance. The non-universal three-loop terms (which are not discussed in the standard bosonization approach) may be different in cases 
$K_1=1/3$, $K_2 =1$ and $K_1=K_2 =1/2$. 


\emph{Conclusion}. In this paper we employed a fermionic description of a
general two-wire junction of two TLL-wires to derive the renormalization
group equation for the conductance, using the approach developed by us
earlier \cite{Aristov2009}. We used an infinite summation of perturbation theory in the
form of a ladder approximation, allowing for an analytical solution for
arbitrary junction parameters and interactions in the wires. As demonstrated
earlier the approximation is asymptotically exact in the vicinity of the
fixed points. As in the well-studied case of a symmetric junction there
exist two fixed points of the RG flow. Fixed point $N$ corresponds to a
complete separation of the wires, i.e. the conductance vanishes. It is
stable in a region of the $K_{1}$-$K_{2}-$coupling constant plane which is
predominantly repulsive, meaning that a weakly attractive component, say 
$K_{2}>1$ is permitted. At fixed point $A$ the conductance assumes its
maximum value. It is stable in the complementary part of the coupling
constant plane. The two stability regions in the $K_{1}$-$K_{2}-$plane are
separated by a hyperbolic boundary curve. The representation chosen is
maximally general and may therefore be easily extended to junctions
connecting more than two wires. The $1$-$2$-symmetric three wire junction 
(``Y-junction'') has been considered by us in Refs. \cite{Aristov2011b,Aristov2011,Aristov2011a,Aristov2010}. Work on the four-wire
junction is in progress.


We thank D.G. Polyakov for helpful comments on the manuscript and A.P. Dmitriev, I.V. Gornyi and V.Yu. Kachorovskii for useful discussions.
The work of D.A. was supported by the German-Israeli-Foundation (GIF), the
Dynasty foundation. The work of D.A. and P.W. was
supported by the DFG-Center for Functional Nanostructures.


\begin{thebibliography}{11}
\expandafter\ifx\csname natexlab\endcsname\relax\def\natexlab#1{#1}\fi
\expandafter\ifx\csname bibnamefont\endcsname\relax
  \def\bibnamefont#1{#1}\fi
\expandafter\ifx\csname bibfnamefont\endcsname\relax
  \def\bibfnamefont#1{#1}\fi
\expandafter\ifx\csname citenamefont\endcsname\relax
  \def\citenamefont#1{#1}\fi
\expandafter\ifx\csname url\endcsname\relax
  \def\url#1{\texttt{#1}}\fi
\expandafter\ifx\csname urlprefix\endcsname\relax\def\urlprefix{URL }\fi
\providecommand{\bibinfo}[2]{#2}
\providecommand{\eprint}[2][]{\url{#2}}

\bibitem[{\citenamefont{Giamarchi}(2003)}]{GiamarchiBook}
\bibinfo{author}{\bibfnamefont{T.}~\bibnamefont{Giamarchi}},
  \emph{\bibinfo{title}{Quantum Physics in One Dimension}}
  (\bibinfo{publisher}{Clarendon Press}, \bibinfo{address}{Oxford},
  \bibinfo{year}{2003}).

\bibitem[{\citenamefont{Gogolin et~al.}(1998)\citenamefont{Gogolin, Nersesyan,
  and Tsvelik}}]{GoNeTs}
\bibinfo{author}{\bibfnamefont{A.~O.} \bibnamefont{Gogolin}},
  \bibinfo{author}{\bibfnamefont{A.~A.} \bibnamefont{Nersesyan}},
  \bibnamefont{and} \bibinfo{author}{\bibfnamefont{A.~M.}
  \bibnamefont{Tsvelik}}, \emph{\bibinfo{title}{Bosonization and Strongly
  Correlated Systems}} (\bibinfo{publisher}{Cambridge University Press},
  \bibinfo{address}{Cambridge}, \bibinfo{year}{1998}).

\bibitem[{\citenamefont{Kane and Fisher}(1992)}]{Kane1992}
\bibinfo{author}{\bibfnamefont{C.~L.} \bibnamefont{Kane}} \bibnamefont{and}
  \bibinfo{author}{\bibfnamefont{M.~P.~A.} \bibnamefont{Fisher}},
  \bibinfo{journal}{Phys. Rev. B} \textbf{\bibinfo{volume}{46}},
  \bibinfo{pages}{15233} (\bibinfo{year}{1992}).

\bibitem[{\citenamefont{Furusaki and Nagaosa}(1993)}]{Furusaki1993}
\bibinfo{author}{\bibfnamefont{A.}~\bibnamefont{Furusaki}} \bibnamefont{and}
  \bibinfo{author}{\bibfnamefont{N.}~\bibnamefont{Nagaosa}},
  \bibinfo{journal}{Phys. Rev. B} \textbf{\bibinfo{volume}{47}},
  \bibinfo{pages}{4631} (\bibinfo{year}{1993}).
  
  \bibitem[{\citenamefont{Safi and Schulz}(1999)}]{Safi1999}
\bibinfo{author}{\bibfnamefont{I.}~\bibnamefont{Safi}} \bibnamefont{and}
  \bibinfo{author}{\bibfnamefont{H.~J.} \bibnamefont{Schulz}},
  \bibinfo{journal}{Phys. Rev. B} \textbf{\bibinfo{volume}{59}},
  \bibinfo{pages}{3040} (\bibinfo{year}{1999}).

\bibitem[{\citenamefont{Weiss et~al.}(1995)\citenamefont{Weiss, Egger, and
  Sassetti}}]{Weiss1995}
\bibinfo{author}{\bibfnamefont{U.}~\bibnamefont{Weiss}},
  \bibinfo{author}{\bibfnamefont{R.}~\bibnamefont{Egger}}, \bibnamefont{and}
  \bibinfo{author}{\bibfnamefont{M.}~\bibnamefont{Sassetti}},
  \bibinfo{journal}{Phys. Rev. B} \textbf{\bibinfo{volume}{52}},
  \bibinfo{pages}{16707} (\bibinfo{year}{1995}).

\bibitem[{\citenamefont{Fendley et~al.}(1995)\citenamefont{Fendley, Ludwig, and
  Saleur}}]{Fendley1995}
\bibinfo{author}{\bibfnamefont{P.}~\bibnamefont{Fendley}},
  \bibinfo{author}{\bibfnamefont{A.~W.~W.} \bibnamefont{Ludwig}},
  \bibnamefont{and} \bibinfo{author}{\bibfnamefont{H.}~\bibnamefont{Saleur}},
  \bibinfo{journal}{Phys. Rev. B} \textbf{\bibinfo{volume}{52}},
  \bibinfo{pages}{8934} (\bibinfo{year}{1995}).

\bibitem[{\citenamefont{Yue et~al.}(1994)\citenamefont{Yue, Glazman, and
  Matveev}}]{Yue1994}
\bibinfo{author}{\bibfnamefont{D.}~\bibnamefont{Yue}},
  \bibinfo{author}{\bibfnamefont{L.~I.} \bibnamefont{Glazman}},
  \bibnamefont{and} \bibinfo{author}{\bibfnamefont{K.~A.}
  \bibnamefont{Matveev}}, \bibinfo{journal}{Phys. Rev. B}
  \textbf{\bibinfo{volume}{49}}, \bibinfo{pages}{1966} (\bibinfo{year}{1994}).

\bibitem[{\citenamefont{Aristov and W\"{o}lfle}(2009)}]{Aristov2009}
\bibinfo{author}{\bibfnamefont{D.~N.} \bibnamefont{Aristov}} \bibnamefont{and}
  \bibinfo{author}{\bibfnamefont{P.}~\bibnamefont{W\"{o}lfle}},
  \bibinfo{journal}{Phys. Rev. B} \textbf{\bibinfo{volume}{80}},
  \bibinfo{eid}{045109} (pages~\bibinfo{numpages}{22}) (\bibinfo{year}{2009}).

\bibitem[{\citenamefont{Aristov et~al.}(2010)\citenamefont{Aristov, Dmitriev,
  Gornyi, Kachorovskii, Polyakov, and W\"olfle}}]{Aristov2010}
\bibinfo{author}{\bibfnamefont{D.~N.} \bibnamefont{Aristov}},
  \bibinfo{author}{\bibfnamefont{A.~P.} \bibnamefont{Dmitriev}},
  \bibinfo{author}{\bibfnamefont{I.~V.} \bibnamefont{Gornyi}},
  \bibinfo{author}{\bibfnamefont{V.~Y.} \bibnamefont{Kachorovskii}},
  \bibinfo{author}{\bibfnamefont{D.~G.} \bibnamefont{Polyakov}},
  \bibnamefont{and} \bibinfo{author}{\bibfnamefont{P.}~\bibnamefont{W\"olfle}},
  \bibinfo{journal}{Phys. Rev. Lett.} \textbf{\bibinfo{volume}{105}},
  \bibinfo{pages}{266404} (\bibinfo{year}{2010}).

\bibitem[{\citenamefont{Aristov and W\"olfle}(2011)}]{Aristov2011a}
\bibinfo{author}{\bibfnamefont{D.~N.} \bibnamefont{Aristov}} \bibnamefont{and}
  \bibinfo{author}{\bibfnamefont{P.}~\bibnamefont{W\"olfle}},
  \bibinfo{journal}{Phys. Rev. B} \textbf{\bibinfo{volume}{84}},
  \bibinfo{pages}{155426} (\bibinfo{year}{2011}).

\bibitem[{\citenamefont{Aristov and W\"olfle}()}]{Aristov2011b}
\bibinfo{author}{\bibfnamefont{D.~N.} \bibnamefont{Aristov}} \bibnamefont{and}
  \bibinfo{author}{\bibfnamefont{P.}~\bibnamefont{W\"olfle}},
  \bibinfo{journal}{arXiv:1110.1159} .


\bibitem[{\citenamefont{Furusaki and Nagaosa}(1996)}]{Furusaki1996}
\bibinfo{author}{\bibfnamefont{A.}~\bibnamefont{Furusaki}} \bibnamefont{and}
  \bibinfo{author}{\bibfnamefont{N.}~\bibnamefont{Nagaosa}},
  \bibinfo{journal}{Phys. Rev. B} \textbf{\bibinfo{volume}{54}},
  \bibinfo{pages}{R5239} (\bibinfo{year}{1996}).
  
  
\bibitem[{\citenamefont{de~C.~Chamon and Fradkin}(1997)}]{Chamon1997}
\bibinfo{author}{\bibfnamefont{C.}~\bibnamefont{de~C.~Chamon}}
 \bibnamefont{and} \bibinfo{author}{\bibfnamefont{E.}~\bibnamefont{Fradkin}},
 \bibinfo{journal}{Phys. Rev. B} \textbf{\bibinfo{volume}{56}},
 \bibinfo{pages}{2012} (\bibinfo{year}{1997}).

\bibitem[{\citenamefont{Aristov}(2011)}]{Aristov2011}
\bibinfo{author}{\bibfnamefont{D.~N.} \bibnamefont{Aristov}},
  \bibinfo{journal}{Phys. Rev. B} \textbf{\bibinfo{volume}{83}},
  \bibinfo{pages}{115446} (\bibinfo{year}{2011}).

\end{thebibliography}

\end{document}